\documentclass[letter]{aa} 

\newcommand{\ic}{IC~1396~N}
\usepackage{graphicx}
\usepackage{txfonts}
\begin{document}
   \title{Protostellar clusters in intermediate mass (IM) star forming regions}
   \author{A. Fuente
          \inst{1}
          \and
          C. Ceccarelli\inst{2}
          \and
          R. Neri\inst{3}
          \and
          T. Alonso-Albi\inst{1}
          \and
          P. Caselli\inst{4,5}
          \and
          D. Johnstone\inst{6,7}
          \and
          E.F. van Dishoeck\inst{8}
          \and
         F. Wyrowski\inst{9}			
          }

   \offprints{A. Fuente}

   \institute{Observatorio Astron\'omico Nacional (OAN), Apdo. 112,
              E-28803 Alcal\'a de Henares (Madrid), Spain \\
              \email{a.fuente@oan.es}
              \and
               Laboratoire d'Astrophysique de l'Observatoire de Grenoble, BP 53, 38041 Grenoble Cedex 9, France
              \and
              Institute de Radioastronomie Millim\'etrique, 300 rue de la Piscine, 38406 St Martin d'Heres Cedex, France
              \and
              INAF-Osservatorio Astrofisico di Arcetri, Largo E. Fermi 5, 50125 Firenze, Italy 
              \and
              Harvard-Smithsonian Center for Astrophysics, 60 Garden Street, Cambridge, MA 0213
              \and
               Department of Physics and Astronomy, University of Victoria, Victoria, BC V8P 1A1, Canada
               \and		
               National Research Council of Canada, Herzberg Institute of Astrophysics, 5071 West Saanich Road, Victoria, BC V9E 2E7, Canada 
              \and
              Leiden Observatory, PO Box 9513, 2300 RA Leiden, Netherlands
             \and
             Max-Planck-Institut fur Radioastronomie, Auf dem Hugel 69, 53121 Bonn, Germany 
}
   \date{Received February 14, 2007; accepted March 29, 2007}

 
  \abstract
  {The transition between the low density groups of T Tauri stars
and the high density clusters around massive stars occurs in the 
intermediate-mass (IM) range (M$_*$$\sim$2--8~M$_\odot$). 
High spatial resolution studies of IM young stellar objects (YSO) 
can provide important clues to understand
the clustering in massive star forming regions.}
  {Our aim is to search for clustering in IM Class 0 protostars.
The
high spatial resolution and sensitivity provided by the new A configuration
of the Plateau de Bure Interferometer (PdBI) allow us
to study the clustering in these nearby objects. }
 {We have imaged three IM Class 0 protostars (Serpens-FIRS~1,
IC~1396~N, CB~3)
in the continuum at 3.3 and 1.3mm using the PdBI. The sources
have been selected with different luminosity
to investigate the dependence of the clustering process
on the luminosity of the source. }
{Only one millimeter (mm) source is detected towards the low luminosity
source Serpens--FIRS~1. Towards CB~3 and IC1396~N, we detect two
compact sources separated by $\sim$0.05~pc. The 1.3mm image of IC~1396~N,
which provides the highest spatial resolution, reveal that one of these
cores is splitted in, at least, three individual sources. }
{}

   \keywords{stars:formation--stars: individual (Serpens--FIRS~1,
IC~1396~N, CB~3)}

   \maketitle
%

\section{Introduction}

Low and high mass stars (M$_*$$>$8~M$_\odot$) are formed in
different regimes. While low mass stars can be formed isolated or 
in loose associations, high mass stars are always found
in tight clusters.
Intermediate-mass young stellar objects (IMs) (protostars and Herbig
Ae/Be [HAEBE] stars with M$_*$ $\sim$ 2 - 8 M$_{\odot}$)
constitute the link between
low- and high-mass stars. In particular the transition between 
the low density groups around T Tauri stars
and the dense clusters around massive stars occurs in these objects. 
Testi et al. (1998,1999) studied the 
clustering around HAEBE stars using optical and near-infrared (NIR) 
images and concluded
that transition occurs smoothly from Ae to Be stars. 
Thus, these stars are key
objects to study the onset of clustering.

Thus far, clustering has only
been studied at infrared and optical wavelengths because of 
the limited 
spatial resolution and sensitivity of the mm
telescopes. Thus, the earliest stages of the cluster
formation were hidden to the observers.
The sub-arcsecond angular resolution provided by the new A configuration
of the PdBI allows, for the first time, to study clustering
at mm wavelengths with a similar sensitivity and spatial 
resolution to the NIR studies.
In this Letter, we present interferometric continuum observations of
the IM protostars Serpens-FIRS 1 (precursor of a Ae star) and CB~3
(precursor of a Be star) aimed to study the clustering phenomena
in the early Class 0 phase. We also use the data
at highest spatial resolution towards \ic\  reported in 
this special issue by Neri et al. (Paper II, hereafter). 

\begin{table*}
\caption{Millimeter flux densities, sizes, spectral indexes and masses}             
\label{table:1}      
\begin{tabular}{l c c c c c c c c c c}     
\hline\hline       
\multicolumn{1}{c}{} & \multicolumn{2}{c}{Position} & \multicolumn{1}{c}{Peak} & 
\multicolumn{1}{c}{Gaussian width$^1$} &  
\multicolumn{1}{c}{Int. Intensity} & 
\multicolumn{1}{c}{Mass$^2$} & \multicolumn{1}{c}{Size$^3$} & 
\multicolumn{1}{c}{$\alpha$$^4$} & \multicolumn{1}{c}{Sensitivity$^5$} &
\multicolumn{1}{c}{Sampled area$^6$} \\
\multicolumn{1}{c}{} & \multicolumn{2}{c}{} & \multicolumn{1}{c}{(mJy/beam)} & 
\multicolumn{1}{c}{($"$)} & \multicolumn{1}{c}{(mJy)} & 
\multicolumn{1}{c}{(M$_\odot$)} & \multicolumn{1}{c}{(AU)} & 
\multicolumn{1}{c}{} & \multicolumn{1}{c}{(M$_\odot$)} &
\multicolumn{1}{c}{r(pc)} 
\\
\hline \hline                    
\multicolumn{7}{l}{Serpens-FIRS~1} \\
1.3mm & 18:29:49.80 & 01:15:20.41 &  273(1)  & 0.50$"$$\times$0.63$"$ & 
357  & 0.1 & 65   & 1.57   & 0.01  & 0.02 \\
3.3mm & 18:29:49.80 & 01:15:20.41 &  63(0.5) & 0.80$"$$\times$1.73$"$ & 71 &
 &   &   &  & 0.04\\
\hline                 
\multicolumn{7}{l}{CB~3-1} \\
1.3mm & 00:28:42.60 & 56:42:01.11 &  20(1)    & 0.36$"$$\times$0.48$"$ & 34  &
0.62 & 600  & 2.52   & 0.04 & 0.16 \\
3.3mm & 00:28:42.60 & 56:42:01.11 &  2.0(0.5) & 0.88$"$$\times$1.27$"$ & 2.9  &
   &  &   & & 0.32 \\
\hline                  
\multicolumn{7}{l}{CB~3-2} \\
1.3mm & 00:28:42.20 & 56:42:05.11 &  10(1)    & 0.31$"$$\times$0.43$"$ & 13  & 
0.24 & 330  & 1.87   &  0.04 & 0.16 \\
3.3mm & 00:28:42.20 & 56:42:05.11 &  2.1(0.5) & 0.80$"$$\times$1.00$"$ & 2.1  & 
& &    &   & 0.32  \\
\hline \hline                  
\end{tabular}

\noindent
$^1$ Half-power width of the fitted 2-D elliptical Gaussian

\noindent
$^2$ Mass estimated using the
1.3mm fluxes and assuming
T$_d$=100~K and $\kappa_{1.3mm}$=0.01~g$^{-1}$~cm$^2$

\noindent
$^3$ Deconvolved source size at 1.3mm

\noindent
$^4$ 1.3mm/3.3mm spectral index

\noindent
$^5$ 5$\times$rms mass sensitivity derived from the 1.3mm image assuming 
T$_d$=100~K and $\kappa_{1.3mm}$=0.01~g$^{-1}$~cm$^2$

\noindent
$^6$ Radius (HPBW/2) of the PdBI primary beam
at the source distance 

\end{table*}

\subsection{Serpens-FIRS~1}  
Serpens-FIRS~1 is a 46~L$_\odot$ Class 0 source located 
in a very active star forming region. 
Previous mid-IR and NIR studies
show that the population of YSOs is strongly clustered, with
the Class I sources more clustered than the Class II ones
(Kaas et al. 2004).
The sub-clusters of Class I sources are located in a NW-SE 
oriented ridge following the 
distribution of dense cores in the molecular cloud
with a subclustering spatial scale of
0.12~pc (see Fig. 1). The Class II stars are located surrounding
the molecular cores with a subclustering spatial
scale of 0.25~pc.  Adopting a distance
of 310~pc, the YSOs density in the sub-clusters ranges 
from 360--780~pc$^{-2}$. Several high angular
resolution mm studies
have been made in the Serpens molecular cloud 
(Testi \& Sargent 1998, William \& Myers
1999, Hogerheijde et al. 1999, Testi et al. 2000). 
We have imaged at higher spatial resolution 
a region of 0.04~pc around the intense mm-source FIRS~1. 

   \begin{figure}
   \centering
    \includegraphics[angle=0,width=8.5cm]{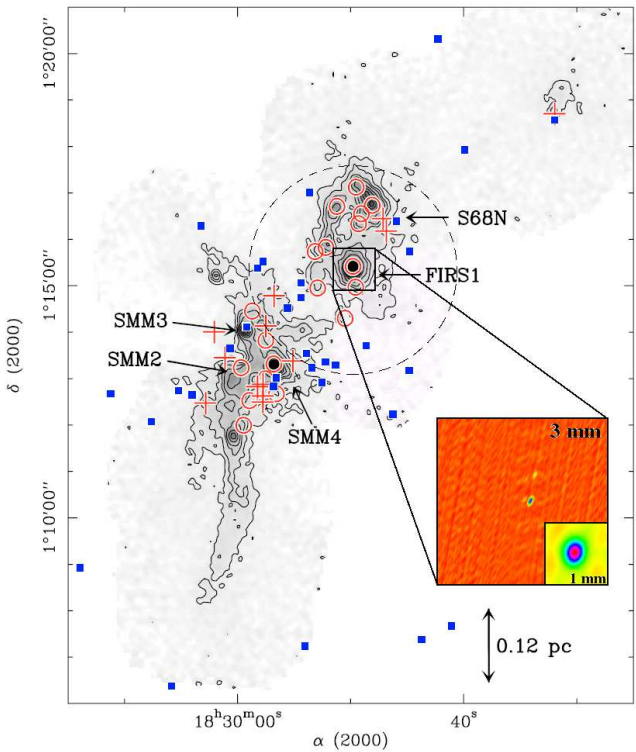}
      \caption{Dust continuum mosaic (contours and grey scale) of the
Serpens main core as observed with the IRAM 30m telescope. The location of
the Class II (blue filled squares), flat (red crosses) and Class I sources
(red empty circles) is indicated (adapted from Kaas et al. 2004). In the inset,
we show the 3mm and 1.3mm (small inset) continuum images observed with 
the PdBI.
Note that only one compact core is detected 
in this region down to a spatial scale of less than 100~AU.
The dashed circle marks a region of 0.2~pc radius around FIRS~1.}
         \label{Fig 1}
  \end{figure}

\subsection{IC~1396~N}
IC~1396~N is a $\sim$300~L$_\odot$ source located
at a distance of 750~pc (Codella et al. 2001).
A total population of $\sim$30 YSOs has been found in 
this region (Getman et al. 2007, Nisini et al. 2001).
These YSOs present an elongated spatial
distribution with an age gradient towards the center
of the Class~I/0 system. The Class III sources are located 
in the outer rim of the globule, the Class II sources are
congregated in the bright ionized rim and the Class I/0
objects are located towards the dense molecular clump
(see Fig. 2). 
The average density
of YSOs in the globule is $\sim$200 pc$^{-2}$. We
have mapped a region of 0.1~pc around the Class 0/I system.
 
\subsection{CB~3}
CB~3 is a large globule (930~L$_\odot$) 
located at 2.5~Kpc from the Sun
(Codella \& Bachiller 1999). A strong submillimeter 
source is observed in the central core (see Fig. 3 and
Huard et al. 2000).
Deep NIR images
of the region show $\sim$40 NIR sources, 
from which at least 22 are very red,
indicative of pre-main sequence stars (Launhardt et al. 1998). 
Up to our knowledge, there are no mid-IR and/or X-ray
studies in this region. Then, the census of YSOs is not
complete in this IM source. We have mapped a
region of 0.32~pc around the submillimeter source. 

   \begin{figure*}
   \centering
   \includegraphics[angle=0,width=18cm]{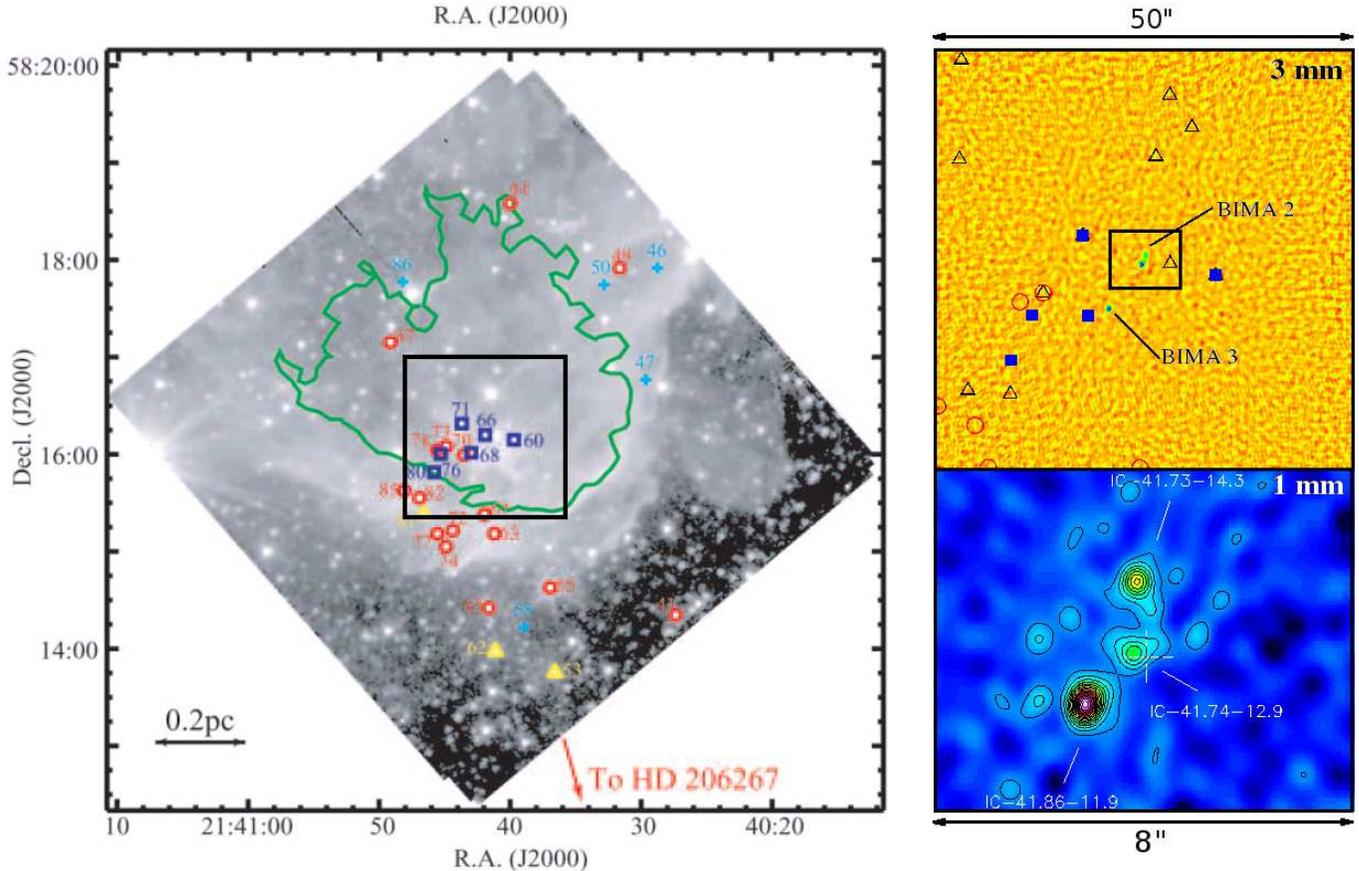}
      \caption{On the left, we show the 5$'$$\times$5$'$ Spitzer IRAC 3.6 $\mu$m image 
towards IC~1396~N (adapted from Getman et al. 2007). The location
of the globule is marked by the green contour and the Class III (yellow
triangles),  Class II (red circles) and blue squares (Class 0/I) sources
are indicated. On the right, we show the 3mm (up) and 1.3mm (down) continuum images 
observed with the PdBI. In the 3mm image we also indicate
the Class III (black triangles),
Class II (red circles) and Class 0/I (filled blue squares) sources. }
         \label{Fig 2}
   \end{figure*}

\section{Observations}
The observations were made on January and 
February, 2006. 
The spectral correlator was
adjusted to cover the entire RF passbands (580\,MHz) for highest
continuum sensitivity. The overall flux scales for each epoch and for
each frequency band were set on 3C454.3 and MWC349 (for CB~3),
and 1749+096 (for Serpens--FIRS~1). The resulting 
continuum point source sensitivities (5$\times$rms) were estimated 
to 2.00\,mJy at 237.571\,GHz 
and 0.5\,mJy at 90.250\,GHz for CB~3 and 40.00\,mJy at 237.571\,GHz 
and 7.0\,mJy at 90.250\,GHz for Serpens--FIRS~1. 
The corresponding synthesized beams adopting uniform
weighting were 0.4$''\times0.3''$ at 237.571\,GHz and
1.0$''\times0.8''$ at 90.250\,GHz for CB~3 and
0.6$''\times0.4''$ at 237.571\,GHz and
1.7$''\times0.7''$ at 90.250\,GHz for Serpens--FIRS~1. 
(See Paper II for IC~1396~N.)

\section{Results}

In Table 1 we present the coordinates, sizes and 
mm fluxes of the compact cores detected
in Serpens--FIRS~1 and CB~3.
The results towards IC~1396~N are presented in
Paper II.
Only 1 mm-source is detected
in Serpens--FIRS~1 
down to a separation of less than 100~AU. 
The other targets turned out
to be multiple sources. We have detected 2 mm-sources 
towards CB~3 and 4 mm-sources towards IC~1396~N.
 
The 4 compact sources towards IC~1396~N are
grouped in 2 sub-clusters
separated by 0.05~pc which are spatially coincident
with the sources named BIMA~2
and BIMA~3 by Beltr\'an et al. (2002). 
The projected distance between these sub-clusters
is similar to that found by Hunter et al. (2007)
between the mm sub-clusters in the massive star forming
region NGC~6336~I. This distance is also similar 
to the distance between the stars forming the Trapezium 
in Orion (from 5000 to 10000 AU). Thus it is a typical distance 
between the IM and massive stars in the same cloud.
Our high angular resolution observations
reveal that BIMA~2 is itself composed of 3 compact cores 
embedded in a more extended component (see Fig. 2).
These 3 compact cores are new mm detections and
constitute the first sub-cluster of Class 0 IM sources
detected thus far.

In CB~3 we have detected 2 mm-sources separated 
by 0.06~pc (see Table 1 and Fig. 3).
These compact cores are new
detections and the separation between them
is similar to that  between BIMA~2 and BIMA~3 
in IC~1396~N.
In fact, the structure of the globule CB~3 resembles
much that of IC~1396~N but the angular resolution of our
observations prevent us from resolving any possible 
sub-cluster of compact cores in this more distant
source. Note that the masses of CB~3-1 and CB~3-2 are 
similar to that
of the sub-cluster BIMA~2 (Paper II).

The number of detections is
limited by the sensitivity of our observations.
In Table 1 we show the point source mass sensitivity
assuming a dust temperature of 100~K (typical for
hot cores and circumstellar disks around 
luminous Be stars) and 
$\kappa_{1.3mm}$=0.01~g$^{-1}$~cm$^{2}$ for each
target. It is possible that we miss a population of
weak Class 0/I sources in CB~3 where the mass sensitivity is 
poor (0.04~M$_\odot$). However, the sensitivity in 
Serpens--FIRS~1 (0.01~M$_\odot$) and IC1396~N 
(0.007~M$_\odot$) is good enough to detect disks around early Be
stars that usually have masses of $\sim$0.01~M$_\odot$ 
(see e.g Fuente et al. 2003, 2006). We should have also detected
massive disks ($\sim$0.1~M$_\odot$) around 
Herbig Ae and T Tauri stars although the dust temperature is 
lower, T$_d$=15--56~K (Natta et al. 2000). 
But there is still the possibility of the existence of 
HAEBE or T Tauri stars with weak circumstellar disks
that are not detected in our mm images.
Another possibility is that we are missing a population of
hot corinos (we refer as $``$hot corino$"$
to the warm material ($\sim$100~K) around a low mass Class 0 protostar
regardless of its chemical composition) 
with masses below the values reported in Table~1. 
Our sensitivity is good enough to detect
a hot corino similar to IRAS~16293--2422 A and B (L$\sim$10~L$_\odot$)
at the distance
of our sources (see Bottinelli et al. 2004). Thus the possible
$``$missed$"$ hot corinos should 
correspond to lower luminosity protostars. Finally,
we can be missing a population of dense
and cold cores. Assuming a dust temperature of 10~K, these 
compact cold cores should have masses of less than 0.17, 0.12 
and 0.7 M$_\odot$ in
Serpens--FIRS~1, IC~1396~N and CB~3 respectively. These masses
are not large enough to form new IM stars.

\section{Discussion}
Testi et al. (1999) studied the clustering around a large sample of
HAEBE stars. In order to quantify the concept, they introduced the
parameter N$_k$, defined as the number of stars in a radius of
0.2~pc, the typical cluster radius.  They showed that rich clusters
are only found around the most massive stars, although the
parameter N$_k$ is highly variable.  Some Be stars are
born quite isolated, while others have N$_k>$70. For our sources this
number is 22 (Launhardt et al. 1998, but the census is not complete),
29 (from Fig. 1) and 28 (Getman et al 2007; Nisini et al. 2001) in
CB3, Serpens and IC~1396~N respectively, where all previously known
YSOs (Class 0, I, II and III) in the regions are considered.

Our maps show 2 sources in CB~3 on a 0.3~pc scale, 1 source in
Serpens-FIRS~1 on a 0.04~pc scale, and 4 sources in IC~1396~N on a 
0.1~pc scale. Defining N$_{mm}$ as the number of mm sources in
a radius of 0.2~pc, we can estimate N$_{mm}$ from
our observations and provide
a revised value for the total number of YSOs at this scale.
In Serpens our interferometric observations do not
add any new mm source to previous data.
We have observed the most intense mm clump in Fig, 1, the most 
likely to be a multiple source, and only found 1 
compact source. 
Based on the 30m map shown in Fig. 1 and assuming that 
all the clumps host only
one source we estimate N$_{mm}$$\sim$7 from a total of 29 YSOs.
In CB~3, our data add 2 new mm sources (N$_{mm}$=2)
to the previous census of YSOs based on NIR studies. 
In IC~1396~N, we estimate N$_{mm}$=4--16.
The upper limit has been calculated assuming a constant
density of mm sources in the region. 
Usually, the Class 0/I stars are not uniformly distributed in the clouds, 
but grouped in sub-clusters that are coincident with the peak of dense cores. 
Thus the value of N$_{mm}$ is very likely close to 4 and we assumed
this number hereafter. Since BIMA~2 and BIMA~3
were previously detected in the X-rays surveys by Getman et al. (2007), 
we only add two new sources (due to the multiplicity of BIMA~2)
to the total number of YSOs in this region.

Summarizing, the total number of YSOs 
is now 29, 24 and 30 for Serpens--FIRS~1,
CB~3 and IC~1396~N respectively. While Serpens--FIRS~1
is an extraordinarily rich cluster 
compared with the clusters around Ae stars 
reported by Testi et al. (1999),
CB~3 and IC~1396~N do not seem 
to become one of the crowded clusters (N$_k$$\sim$70) detected by 
these authors around Be stars.
However, this conclusion might not be true. 
The interferometer is only sensitive to dense and compact cores
and provides a biased vision of the star forming
regions. In fact our interferometric observations accounts for
less than 1\% of  the total interstellar mass in the studied globules, i.e., 
$\sim$ 10, 58 and 64 M$_\odot$ are missed in Serpens--FIRS~1, 
CB~3 and IC~1396~N respectively
(Alonso-Albi et al. 2007). One possibility is that
this mass is in the form of many weak hot corinos which could
eventually become low mass stars. The fate
of these hot corinos is, however, linked to the evolution of the IM protostar
that is progressively dispersing and warming the surrounding material
(Fuente et al. 1998).
Another possibility is that the $"$missed$"$ mass is in the form of an 
extended and massive envelope. This envelope (if not totally dispersed by 
the IM star) could produce new stars in a forthcoming star formation
event. 

   \begin{figure}
   \centering
    \includegraphics[angle=0,width=8.5cm]{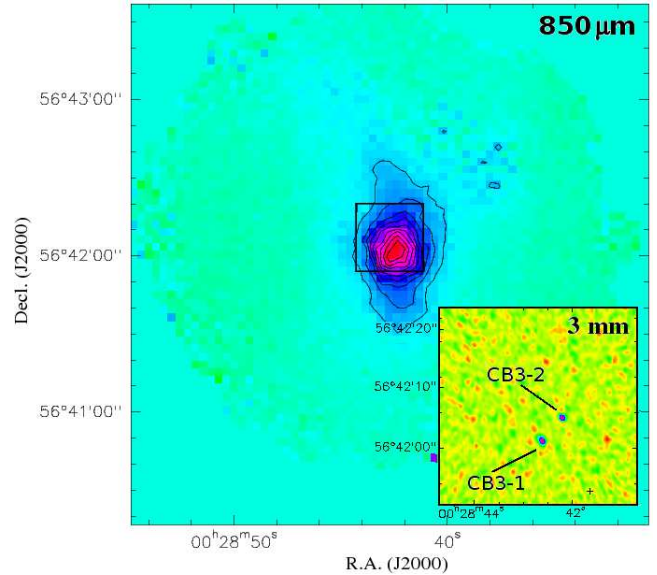}
      \caption{Dust continuum emission at 850~$\mu$m as observed with
SCUBA towards CB~3. In the inset,
we show the 3mm continuum image observed with PdBI. 
Note that two compact cores are detected 
towards the single-dish peak.}
         \label{Fig 1}
  \end{figure}

\section{Summary}
We have searched for clustering at mm wavelengths in
3 IM star forming regions. We have detected
1, 2 and 4 compact cores in Serpens--FIRS~1, CB~3 and IC~1396~N
respectively. The compact cores are not distributed uniformly
but grouped in sub-clusters separated by $\sim$0.05~pc.
Such a separation is a typical distance for both IM and massive
stars within the same cloud. We have used our mm observations
to complete the census of YSOs in these regions and compare
them with the clusters found by Testi et al. (1999) in the
more evolved HAEBE stars. Serpens--FIRS~1 seems to belong to 
an extraordinarily rich cluster. The density of
YSOs in the high luminosity sources IC~1396~N and CB~3 is consistent
with the density found in the clusters around Be stars although
our sources are not found between the most crowded regions.
The large amount of interstellar gas and dust in the studied
regions suggest that new star formation events are still possible.

\begin{acknowledgement}
We are grateful to Phil Myers for his careful reading of the manuscript.
A.F.\ is grateful for support from the Spanish MEC and FEDER funds under grant 
ESP~2003-04957, and from SEPCT/MEC under grant AYA 2003-07584.
\end{acknowledgement}

\end{document}